# STARTUP & UNICORN GROWTH VALUATION

By Dr Andreas A. Aigner & Walter Schrabmair

*How do you value companies which have IPOed recently? How do you compare them amongst their peers? Valuing companies using a linear extrapolation of their revenues and profits leads to an ingenious method to benchmark stocks against each other. Here we present such a method, dubbed the growth average U1.*

> Be a unicorn in a field of horses.

## SECTION 1 – INTRODUCTION

Valuing a startup and young company is often difficult. Especially if they are expanding so fast such that their cash burn is faster than their profits. This is usually the case for SaaS (Software as a Service) or NaaS (Network as a Service) companies that prioritize their expansion over the initial profits, usually offering their services or products at below cost prices in order to grow their customer base and in order to displace any pre-existing competitors out of the market or grab market share from them. This can not only be said about SaaS/NaaS firms but essentially applies to any company trying to establish themselves in the market place, take for example Amazon, Netflix etc.

When the stock has IPOed you can look at snapshots of their ratios such as Price to Sales (P2S) or Price to Earnings (PE) Ratios (provided earnings are positive), and if the company hasn't IPOed you will look at revenue, sales, number of customers. In particular, you would want to measure the growth of these figures in order to predict what a future value of the firm would be. You would infer from the revenue a market capitalization comparing them to other companies in the sector or industry with the average or maximum P2S (or PE).

One such method is the 'Rule of 40' [1, 2] which states that growth rate and profit margin together should exceed 40%. It is used as a quick method to gauge if a stock is attractive or not. Jim Cramer [3] suggested a modified version even that requires this percentage amount to be more than 5 times larger than its P2S.

This Rule of 40 is essentially a sum of two growth percentages. The revenue growth and the profit margin. The profit margin is the ratio of profit over revenues. It isn't quite comparable in nature to the revenue growth since the profit margin is essentially an amount at the product level and increases by multiples of the number of products sold. A company with a small profit margin but a huge number of sales can grow faster than a company with a much bigger profit margin but a small number of sales. This Rule of 40 therefore is not perfect. But it is a quick and dirty way to come up with a mechanical rule to compare companies where you have negative earnings and only a few data points to work with.

# SECTION 2 – AVERAGE GROWTH = U1

Here we aim to come up with a better, more accurate method for valuing such companies. When you follow stocks that have IPOed only recently and that have a strong price trend they all have something in common when you look at their financials (See Figure 1 Typical Revenue and Earnings). They have a revenue growth that looks extraordinarily strong and often times an earnings growth that is a small negative at best (i) or declining steadily/rapidly at worst (ii and iii). From looking at the earnings and revenue rising/declining, what is the best way to gauge if it's a healthy or weak growth. Similar to the Rule of 40 you would calculate what the growth of revenue is.

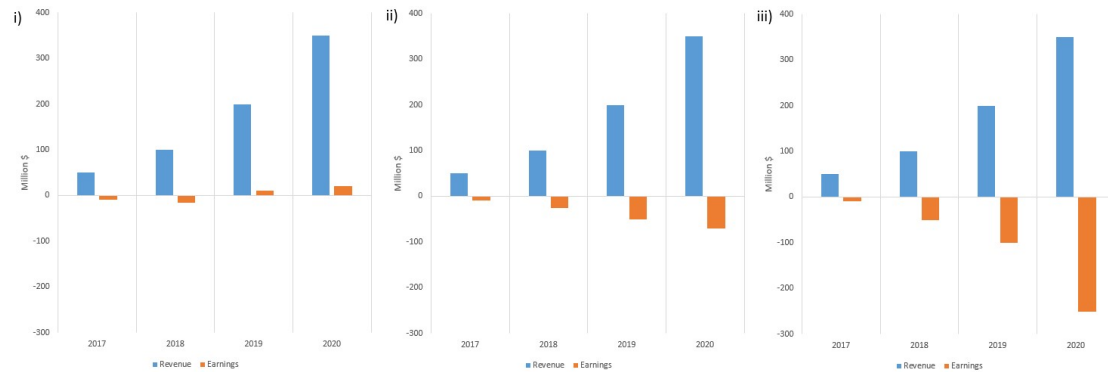

Figure 1 Typical Revenue and Earnings

A perfect way to track the total revenue and gross profit is to use the trailing twelve month running values. Using a trailing twelve-month figure eliminates any seasonality effects which is an advantage already. Looking at an example of these numbers for a stock (CRWD - Crowdstrike) in Figure 2: Total Revenue (TR) and Gross Profit (GP) of CRWD below you notice that the total revenue gets updated at more or less quarterly values and the gross profit every six months. This might vary from stock to stock or release to release. The jump from one level to the next is the amount of growth they have achieved in this period. How do you calculate the annual growth estimate from that? The best way to do this is to calculate a linear regression through the data points, which is indicated by the blue and red dotted lines in Figure 2. Linear regression optimizes the error around fitting a line through the data points. As a result, we get a slope 'm' and a constant 'b', representing the intersection with the y-axis, which is all is needed to define a line through the data points. From the definition of this line it is trivial to extrapolate it out one year and derive the annual % growth value of total revenue and gross profit likewise.

Formulas for 'm' and 'b' can be found in any undergraduate math book, but a readymade function to calculate this directly in a database using standard commands does not exist. But as it turns out it is doable directly in the database without having to outsource this operation to another programming language. We have listed these neat lines of our code for calculating the linear extrapolation in Appendix 1 since we believe users will find this useful in the future.

There is also the question of how long the time window is that you will use to calculate the linear regression. For relatively young companies or startups a period of a maximum of 1-year will be enough to determine the growth. In any case using the twelve month running totals means with a 1-year range you are effectively covering a 2-year period already. Why don't we use longer time windows? Essentially using a much longer time window means we are looking at a long term growth of a seasoned company. We are going to assume that such growth will be already priced into the stock and what we are looking for is shorter term deviations of price and value. This is why we pick a 250 day period, a bit less than a calendar year.

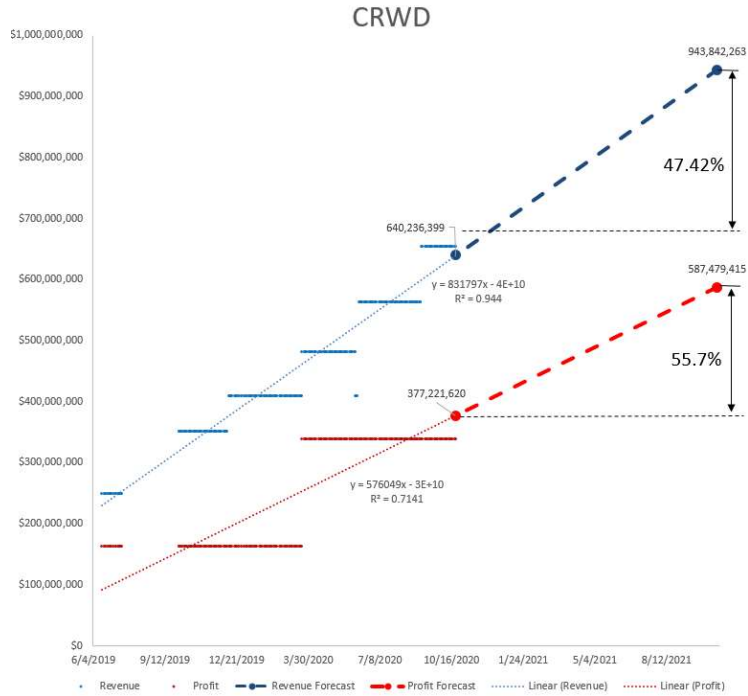

Figure 2: Total Revenue (TR) and Gross Profit (GP) of CRWD

For our example stock CRWD this is a 47.42% growth of total revenue and 55.7% growth of gross profits. We are going to look at other example stocks to elaborate on this. Figure X2 shows the same chart for FVRR, TSLA, AAPL and FB. While stocks like CRWD and FVRR are examples of startups and possible 'unicorn' candidates, it can be argued that stocks like TSLA, AAPL and FB do not fit this moniker anymore. But judging from the growth figures we arrive at interesting conclusions for these 5 stocks.

Table 1 Example Growth Figures shows the derived growth values for total revenue and gross profits and an average of these two. What is noticeable from these example names is that the growth of revenue and growth of profits is of a similar order of magnitude. The gross profits are all positive in these examples also. In general, this might not be the case, there will be companies which have negative profits early on in their history and survive even with negative earnings if they find loans to keep their business afloat. In addition, their revenue growth might be outsizing their decline or growing increase in losses. The growth of gross profits might be negative but the amount of positive growth outweighing the negative growth in earnings. For many investors an increasing loss in earnings is a negative sign, but for some investors it is the net growth of revenue over earnings that counts. They discount the increasing loss in earnings against the outsized growth in revenue. When we calculate the growth values individually we can add the two together in the end and divide by 2, arriving at the average growth. If for example the growth of gross profits (GP) is -20%, but the growth of total revenue (TR) is +50%. The net average

growth would be 15%. This is a pragmatic and quite reasonable way to compare the growth values to other companies doing this.

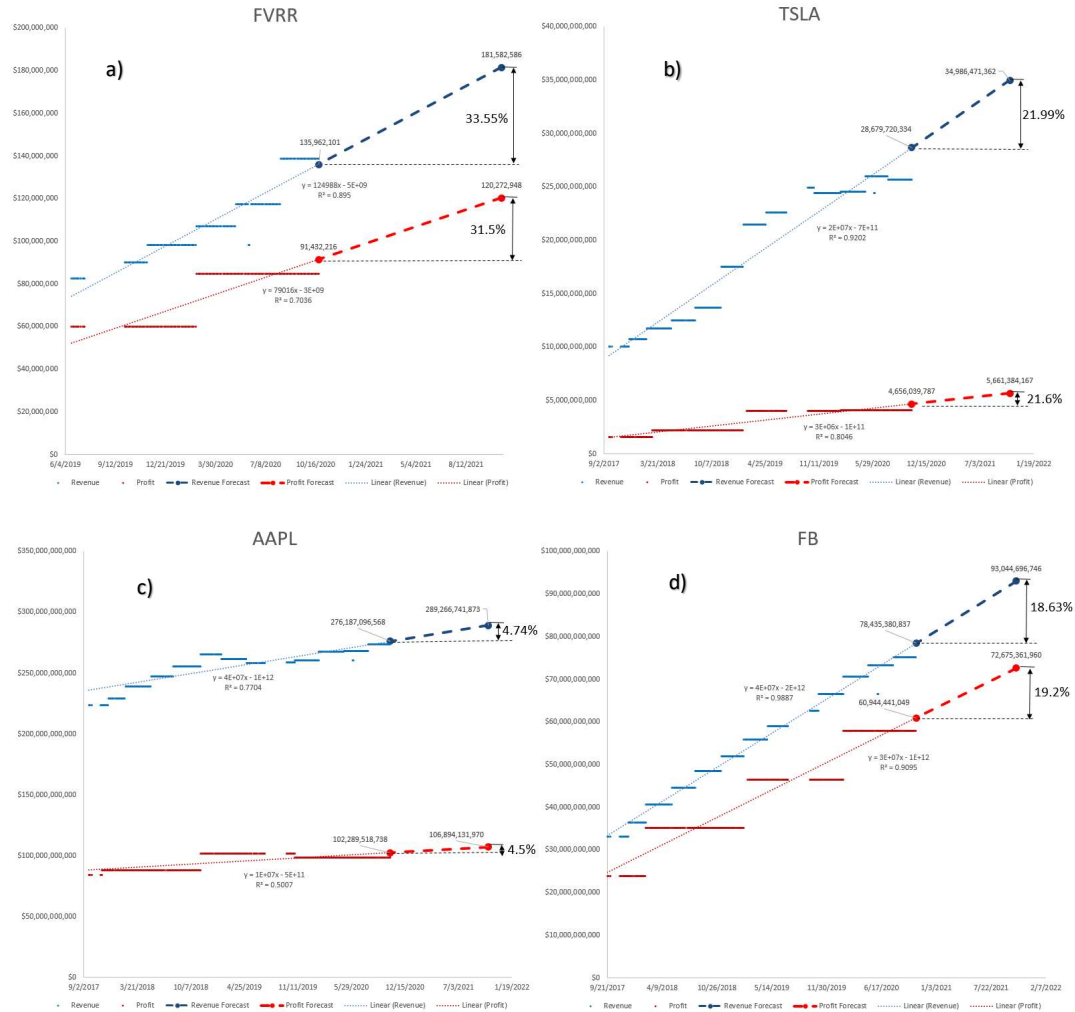

Figure 3: Total Revenue (TR) and Gross Profit (GP) of (a) FVRR, (b) TSLA, (c) AAPL and (d) FB

|      | Total Revenue Growth | Gross Profit Growth | Average Growth |
|-----:|---------------------|---------------------|----------------|
| CRWD | 47.42%              | 55.7%               | 51.56%         |
| FVRR | 33.55%              | 31.5%               | 32.53%         |
| TSLA | 21.99%              | 21.6%               | 21.80%         |
| AAPL | 4.74%               | 4.5%                | 4.62%          |
| FB   | 18.63%              | 19.2%               | 18.92%         |

Table 1 Example Growth Figures

Looking at Table 1 Example Growth Figures you see that MegaCaps like AAPL have a much lower average growth (5%) than young startups like CRWD or FVRR (30-50%), but other MegaCaps like TSLA or FB offer better growth than AAPL does. Judging from these growth estimates they offer a valuable insight into the ranking of stocks in terms of future performance. But there is one aspect still missing from all of this. Price.

## SECTION 3 – U1 CONSTRAINTS

While we have derived values for the growth of companies to compare we have not got a measure of determining price so far. Benjamin Graham's formula for growth stocks does something similar [4], it takes an estimate for a minimum Price to Earnings ratio adds to that a growth value with a factor and multiplies out with the earnings value to arrive at a price estimate. But Graham's Formula only works for positive earnings. If you have very little data available and earnings are negative you can use another common ratio used often to gauge the value of a company, price to sales (P2S). It is the ratio of MarketCap (Market Capitalization) to Revenue.

Comparing the P2S of such growth stocks with others gives us an idea of how expensive the stock is compared to its peers or stocks in the same market cap or other benchmark stocks in general. We can also look at the absolute value of the P2S. When a company's MarketCap is at 5 times its revenue, meaning that 5 years of its revenue sums up to the total market cap, we may label such a stock as 'fairly' priced or even 'cheap'. While if we are looking at a company which is 50 times its revenue, you would need to wait 50 years for its revenue to equal its market cap, which is 'overpriced' or even 'ridiculous', since we are assuming its revenue its equal to its profits, which it is not obviously and many other factors would need to be deducted from the revenue to arrive at the actual 'value' of such a company.

What does the distribution of P2S values look like across a typical set of stocks? Figure X shows a histogram of P2S values across all stocks with a market cap of at least 10bn$ and a total revenue of at least 10mio$ (we filter out the smaller companies for simplicity since they can be quite noisy and don't change the point we are making)

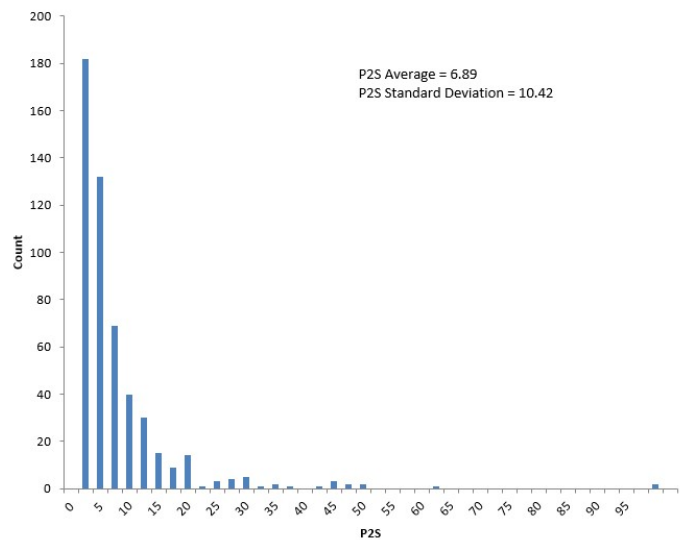

Figure 4: Histogram of P2S

As we notice from the histogram most of the stocks are below 20 P2S. In fact, the P2S average is 6.89 and the standard deviation σ is 10.42. That means that 95% of stocks are within 27.73 P2S. We see that P2S values >30 are rare and values of >50 are extreme outliers being more than 4 standard deviations away from the mean.

(1) We will restrict our growth candidates to a P2S < 2σ (27.73).

In addition to the P2S requirement there are a couple of other requirements that we want to impose on our group of growth candidates.

(2) We want to require that there is at least 1 analyst following the stock. The analyst might have not updated his coverage in a while, but at least there is still a current analyst who has covered the stock in the last 12 months.

(3) We also want to filter for stocks with a positive total revenue. It might not seem obvious but we want to make sure that the business is producing revenue and not simply burning through borrowed cash.

(4) In addition to the TR being greater zero, we also want to make sure that if the gross profits are negative that they are not bigger than 25% of total revenue. Which means that we will allow for a loss in profits as long as they don't exceed ¼ of the total revenue. This puts our growth stocks in a ballpark where either total revenue and gross profits are both positive, or where the total revenue is positive but the gross profits are a small percentage loss of the total revenue.

(5) One additional constraint we want to use is a trading parameter, the Sharpe Ratio, which is associated with our trading algorithm used. It is a trend following strategy according to Wilders' Volatility Index method described here [5-7]. It measures how well the simple trend following strategy works and indirectly measures the 'trend'. A simple buy and hold strategy will typically yield around 0.75 and better trending names will be a lot higher (2-4), so it is only a very weak constraint to require this Sharpe Ratio to be >0 for our group of growth stocks. It in effect just filters out names which are too illiquid or too volatile and just too dangerous to trade.

All the 5 requirements are summarized in the list below.

*Requirements*

1) P2S < 2σ (27.73)
2) Number of Analysts >= 1
3) Total Revenue > 0
4) Gross Profit, ie. Loss no greater than 25% of Total Revenue. $\frac{TR}{GP} > 0$ or $\frac{TR}{GP} < -4$.
5) Sharpe Ratio > 0

Considering the restrictions on total revenue, gross profit and number of analysts, what does a distribution of P2S versus our growth average estimate look like? Looking at Figure 5 you see the distribution of P2S versus U1 for all the MarketCap groups. What one can notice by visual inspection is that the distribution along U1 has a concentration around 0-20 but that the distribution along P2S is not really giving us any insights. But when one inspects the relationship of P2S with U1 it appears almost as if the data points are sloping upwards with increasing U1. It suggests that there is a parabolic frontier where most stocks are below: stocks P2S increases with increasing U1. But there is not really enough statistical evidence to suggest a strong relationship. A reason could be that a stock doesn't have to have an increasing P2S that exceeds its average growth, it could just grow in line with it. But nevertheless this is an interesting chart to inspect. We have highlighted the area of stocks we are looking for in this chart and the red line represents the P2S level of 2σ.

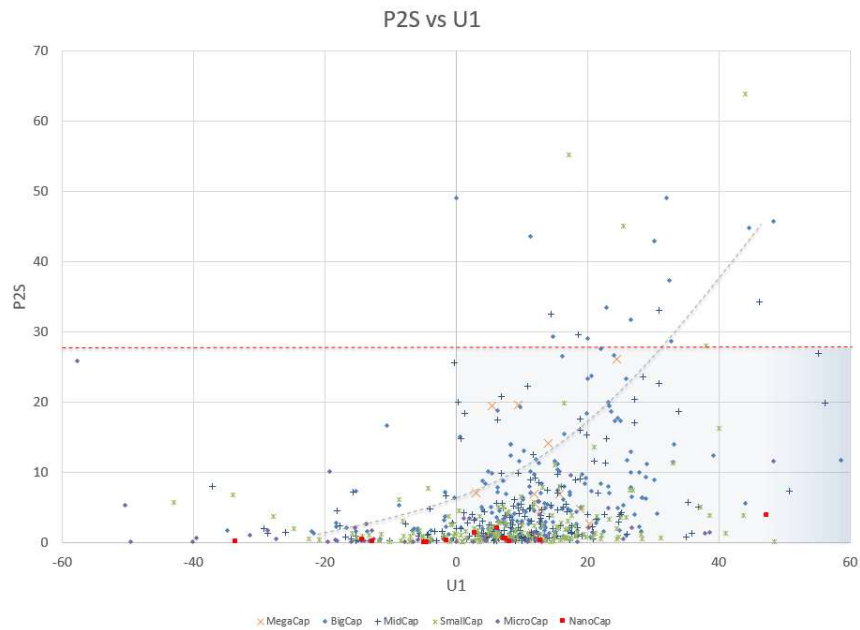

Figure 5: Price-to-Sales Ratio (P2S) versus the average growth (U1)

Figure 6 shows a plot of Sharpe Ratio versus U1. There is no clear relationship between the Sharpe Ratio and the U1 measure, this could be due to several reasons. While the Sharpe Ratio of our trend following algorithm would represent a measure of how well we can follow the trend, it fluctuates with secular market trends as well. But we don't want to pick stocks which are worse than a regular buy-n-hold strategy (0.75) so we at least put a constraint of >0 in place, this is the highlighted area in Figure 6.

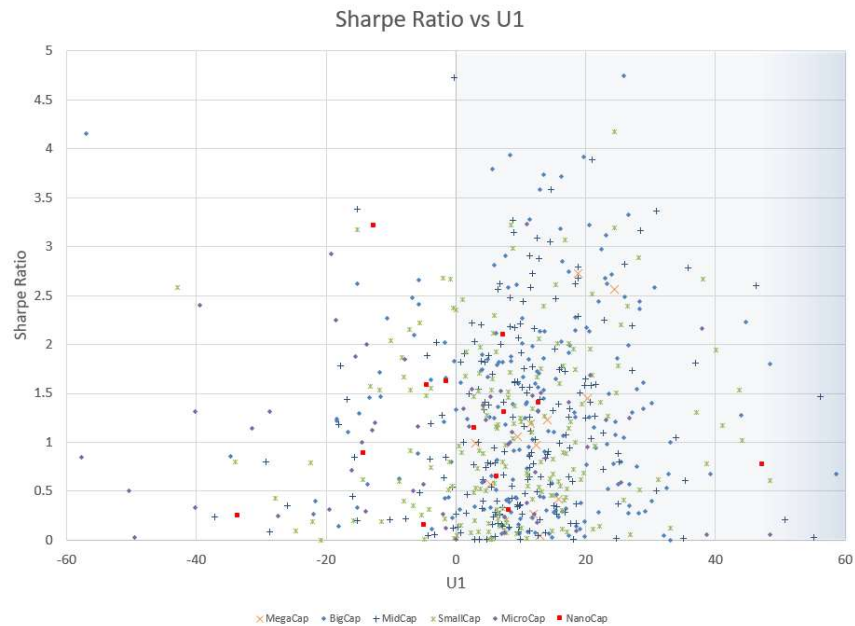

Figure 6: Sharpe Ratio versus average growth U1

Finally we want to look at the relationship of P2S to Sharpe Ratio in Figure 7. One could have expected to see higher Sharpe Ratios with increasing P2S. But there is no apparent relationship and it shows just how most P2S are below 20 and that Sharpe Ratios are also mostly below 3.

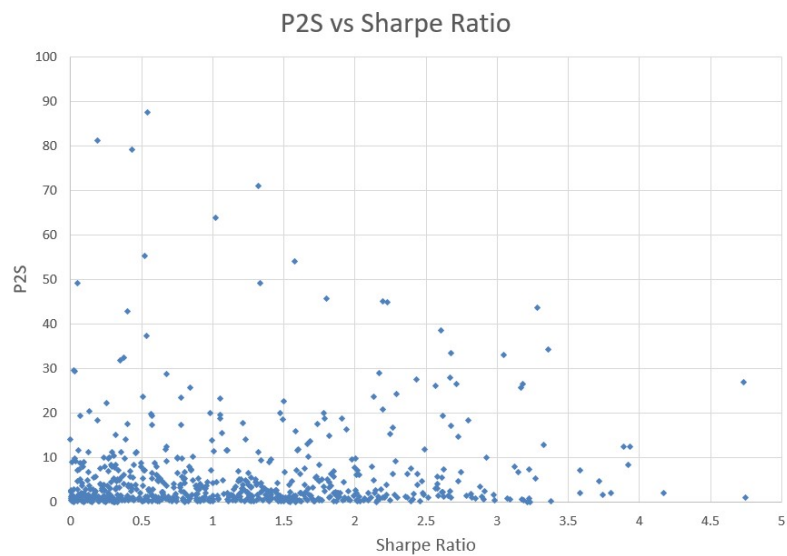

Figure 7: Price-to-Sales Ratio (P2S) versus Sharpe Ratio

# SECTION 4 – CONCLUSION

In Section 2 we have presented a measure to value a stock with minimal financial data. This being its main strength: it simply needs two balance sheet figures to work with, total revenue and gross profits. These are figures that should be available for any business that is running and fundamental. A linear extrapolation of the history of the trailing twelve-month total revenue and gross profits is done to derive the average annual growth value, U1. Unlike traditional valuation methods, we don't need to value the price of the stock. Instead we simply compare this average growth value U1 across all names. In order to filter out unattractive candidates we introduce a list of 5 constraints in SECTION 3 – U1 Constraints. These consist of a restriction to 95% or (2s) of the population in terms of Price to Sales (P2S), a positive total revenue and Sharpe Ratio, a minimum coverage of 1 analyst and positive gross profits or a maximum of 25% of total revenue loss.

Appendix 2 lists the Top10 average growth values U1 for all market capitalizations (MegaCap, BigCap, MidCap, SmallCap, MicroCap, NanoCap)

One obvious criticism or improvement that can be made is by investigating if there is some evidence that links a selection methodology like this to generating profits. We have only looked at how it relates to a basic trend following strategy and have found no relationship. But traditional valuation methods and rule of thumbs like the Rule-of-40 do suggest that there has to be some relationship.

When we look at stocks like CRWD or FVRR in particular it is apparent from the price history that there are points in its price history where it was trading at levels below 20 P2S for quite some time, yet their revenue and profits projections would have already been showing a strong growth. When you have the P2S shooting up before the growth in revenue or profits is measurable then you will miss it. But it seems it is possible to find stocks where the opposite happens. The key to this strategy therefore is to discover these stocks early on and pick those with a good growth outlook. One has to still exercise some judgement in what is a strong growth candidate but that is the case with any investment or trading methodology. At monitoring list like this will give us a look at the best candidates when a more general market event leads to cheap valuation levels in a downturn for example and the market becomes oversold.

# APPENDIX 1 – MYSQL – LINEAR EXTRPOLATION CODE

```sql
SLOPE_GP*31622400/(SLOPE_GP*UNIX_TIMESTAMP()+INTERCEPT_GP AS GROWTH_GP,
SLOPE_TR*31622400/(SLOPE_TR*UNIX_TIMESTAMP()+INTERCEPT_TR AS GROWTH_TR
FROM (SELECT *,
    (SUM_Y_GP) - SLOPE_GP * SUM_X)/NN AS INTERCEPT_GP,
    (SUM_Y_TR) - SLOPE_TR*SUM_X)/NN AS INTERCEPT_TR
FROM (SELECT *,
    (NN*SUM_XY_GP - SUM_X * SUM_Y_GP)/(NN*SUM_XX - SUM_X*SUM_X) AS SLOPE_GP,
    (NN*SUM_XY_TR - SUM_X*SUM_Y_TR)/(NN*SUM_XX-SUM_X*SUM_X) AS SLOPE_TR
FROM (SELECT COUNT(UNIX_TIMESTAMP(DATE)) AS NN,
            SUM(UNIX_TIMESTAMP(DATE)) AS SUM_X,
            SUM(GROSS_PROFITS) AS SUM_Y_GP,
            SUM(TOTAL_REVENUE) AS SUM_Y_TR
            SUM(GROSS_PROFITS*DATE) AS SUM_XY_GP,
            SUM(TOTAL_REVENUE*DATE) AS SUM_XY_TR,
            SUM(DATE*DATE) AS SUM_XX
            FROM DATA WHERE DATE >= (CURDATE() - INTERVAL 250 DAY)
GROUP BY SYMBOL) AS AA) AS BB) AS CC
```

```
*DATE is the UNIX timestamp history array.
** GROSS_PROFITS & TOTAL_REVENUE is the GP and TR history array.
```

# APPENDIX 2 – TOP 10 U1 BY MARKET CAPITALIZATION

**Table 2: Top10 U1 MegaCaps**

| Symbol | U1 | P2S | Sharpe Ratio | TR (Mio) | GP (Mio) | Ratio | P2S Fcst |
|---|---|---|---|---|---|---|---|
| NVDA | 24.48 | 26.09 | 2.57 | 13065 | 6768 | 1.93 | 20.96 |
| HD | 20.24 | 2.67 | 1.45 | 119318 | 37572 | 3.18 | 2.22 |
| PG | 18.89 | 4.81 | 2.73 | 72470 | 36314 | 2.00 | 4.05 |
| GOOGL | 15.8 | 6.89 | 0.42 | 171704 | 89961 | 1.91 | 5.95 |
| TSLA | 14.07 | 14.16 | 1.23 | 28176 | 4069 | 6.92 | 12.41 |
| AMZN | 12.94 | 4.67 | 0.05 | 347945 | 114986 | 3.03 | 4.13 |
| WMT | 12.32 | 0.74 | 0.98 | 542026 | 129359 | 4.19 | 0.66 |
| GOOG | 11.92 | 6.88 | 0.27 | 171704 | 89961 | 1.91 | 6.15 |
| UNH | 11.54 | 1.33 | 1.19 | 252575 | 57598 | 4.39 | 1.19 |
| MA | 9.43 | 19.64 | 1.06 | 15595 | 16883 | 0.92 | 17.95 |
| V | 5.4 | 19.46 | 0.57 | 21846 | 21119 | 1.03 | 18.46 |
| AAPL | 2.96 | 7.12 | 0.99 | 274515 | 104956 | 2.62 | 6.92 |

**Table 3: Top10 U1 BigCaps**

| Symbol | U1 | P2S | Sharpe Ratio | TR (Mio) | GP (Mio) | Ratio | P2S Fcst |
|---|---|---|---|---|---|---|---|
| KKR | 1336.54 | 6.69 | 1.51 | 4695 | 7907 | 0.59 | 0.47 |
| MTCH | 112.96 | 6.84 | 0.13 | 4895 | 3634 | 1.35 | 3.21 |
| CRM | 58.67 | 11.76 | 0.67 | 19380 | 12863 | 1.51 | 7.41 |
| TTWO | 43.98 | 5.54 | 1.27 | 3380 | 1547 | 2.18 | 3.85 |
| ETSY | 39.17 | 12.46 | 0.68 | 1378 | 547 | 2.52 | 8.95 |
| EXAS | 33.11 | 14.02 | 0.00 | 1321 | 660 | 2.00 | 10.53 |
| TAL | 33.01 | 11.51 | 1.00 | 3701 | 1805 | 2.05 | 8.65 |
| LOGI | 30.67 | 4.09 | 2.58 | 3661 | 1137 | 3.22 | 3.13 |
| FTCH | 30.2 | 8.93 | 1.40 | 1334 | 460 | 2.90 | 6.86 |
| QDEL | 29.1 | 11.15 | 0.29 | 1005 | 321 | 3.13 | 8.64 |

**Table 4: Top10 U1 MidCaps**

| Symbol | U1 | P2S | Sharpe Ratio | TR (Mio) | GP (Mio) | Ratio | P2S Fcst |
|---|---|---|---|---|---|---|---|
| ESTC | 56.16 | 20.07 | 1.47 | 467 | 305 | 1.53 | 12.85 |
| GWPH | 50.77 | 7.52 | 0.21 | 442 | 284 | 1.56 | 4.99 |
| DQ | 36.94 | 5.14 | 1.81 | 505 | 80 | 6.31 | 3.75 |
| PFSI | 35.84 | 1.35 | 2.78 | 2982 | 1778 | 1.68 | 0.99 |
| MIME | 35.36 | 5.88 | 0.61 | 462 | 318 | 1.45 | 4.34 |
| CLF | 35.05 | 0.85 | 0.01 | 3632 | 576 | 6.31 | 0.63 |
| NKTR | 33.95 | 18.74 | 1.05 | 162 | 93 | 1.74 | 13.99 |
| FOLD | 30.87 | 23.64 | 0.51 | 227 | 160 | 1.42 | 18.06 |
| PLUG | 28.4 | 25.67 | 3.17 | 260 | 28 | 9.29 | 19.99 |
| INSM | 27.19 | 20.82 | 2.19 | 164 | 112 | 1.46 | 16.37 |

**Table 5: Top10 U1 SmallCaps**

| Symbol | U1 | P2S | Sharpe Ratio | TR (Mio) | GP (Mio) | Ratio | P2S Fcst |
|---|---|---|---|---|---|---|---|
| ORC | 76.78 | 24.33 | 2.29 | 15 | 32 | 0.47 | 13.76 |
| AKBA | 63.89 | 1.08 | 0.65 | 340 | 27 | 12.59 | 0.66 |
| TPC | 48.41 | 0.14 | 0.61 | 4894 | 409 | 11.97 | 0.09 |
| NEWT | 43.74 | 3.8 | 1.54 | 94 | 59 | 1.59 | 2.64 |
| PMT | 41.04 | 1.3 | 1.18 | 1203 | 645 | 1.87 | 0.92 |
| CELH | 40.01 | 16.33 | 1.94 | 103 | 31 | 3.32 | 11.66 |
| CTMX | 38.64 | 3.85 | 0.78 | 85 | 57 | 1.49 | 2.78 |
| AVAV | 37.11 | 5.04 | 1.30 | 368 | 153 | 2.41 | 3.68 |
| CTSO | 33.1 | 11.31 | 0.12 | 32 | 18 | 1.78 | 8.50 |
| ICHR | 31.13 | 0.7 | 0.62 | 859 | 86 | 9.99 | 0.53 |

**Table 6: Top10 U1 MicroCaps**

| Symbol | U1 | P2S | Sharpe Ratio | TR (Mio) | GP (Mio) | Ratio | P2S Fcst |
|---|---|---|---|---|---|---|---|
| **MIXT** | 3279.43 | 1.34 | 0.01 | 131 | 93 | 1.41 | 0.04 |
| **CLSD** | 73.65 | 9.14 | 0.68 | 7 | 2 | 3.50 | 5.26 |
| **LIVX** | 65.97 | 3.72 | 0.67 | 40 | 6 | 6.67 | 2.24 |
| **CLXT** | 48.36 | 11.55 | 0.06 | 11 | -2 | (5.50) | 7.79 |
| **ITI** | 38.62 | 1.42 | 0.06 | 117 | 48 | 2.44 | 1.02 |
| **LTRX** | 26.68 | 2.08 | 1.25 | 60 | 27 | 2.22 | 1.64 |
| **SBBP** | 23.14 | 4.96 | 1.23 | 28 | 18 | 1.56 | 4.03 |
| **MTBC** | 20.72 | 1.41 | 1.69 | 74 | 23 | 3.22 | 1.17 |
| **CWCO** | 20.21 | 2.14 | 0.95 | 73 | 26 | 2.81 | 1.78 |
| **VSTM** | 17.4 | 9.55 | 1.41 | 22 | 16 | 1.38 | 8.13 |

**Table 7: Top10 U1 NanoCaps**

| Symbol | U1 | P2S | Sharpe Ratio | TR (Mio) | GP (Mio) | Ratio | P2S Fcst |
|---|---|---|---|---|---|---|---|
| **KMPH** | 47.28 | 3.83 | 0.77 | 12 | 10 | 1.20 | 2.60 |
| **HDSN** | 12.82 | 0.29 | 1.40 | 155 | 17 | 9.12 | 0.26 |
| **TLGT** | 8.14 | 0.05 | 0.31 | 55 | 13 | 4.23 | 0.05 |
| **REPH** | 7.53 | 0.45 | 1.31 | 80 | 48 | 1.67 | 0.42 |
| **NETE** | 7.28 | 0.48 | 2.10 | 63 | 10 | 6.30 | 0.45 |
| **IZEA** | 6.3 | 1.94 | 0.65 | 18 | 10 | 1.80 | 1.83 |
| **HCAP** | 2.86 | 1.38 | 1.14 | 13 | 13 | 1.00 | 1.34 |
| **REI** | -1.47 | 0.26 | 1.62 | 146 | 144 | 1.01 | 0.26 |
| **EXPR** | -4.46 | 0.03 | 1.58 | 1551 | 901 | 1.72 | 0.03 |
| **CBL** | -4.9 | 0.03 | 0.15 | 669 | 538 | 1.24 | 0.03 |